\documentclass[12pt]{article}

\usepackage{amsfonts}
\usepackage{amsmath}
\usepackage[position=top]{subfig}
\usepackage{graphicx}
\usepackage{hyperref}

\def\R{\mathbb{R}}
\def\P{\mathbb{P}}
\def\N{\mathbb{N}}

\title{\sc Population genetics: coalescence rate and demographic parameters inference}
\author{Olivier Mazet\footnote{Institut de Mathématique de Toulouse (UMR 5219) and Institut National des Sciences Appliquées de Toulouse} ~ and Camille Noûs\footnote{Laboratoire Cogitamus}}
\date{May 2, 2023}

\begin{document}
\maketitle
\vspace{-1cm}
\begin{figure}[ht]
 \centering
 \includegraphics[width=0.25\textwidth]{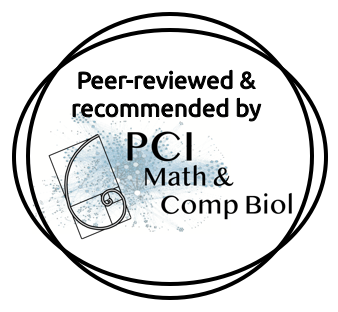}
 \caption*{
\url{https://mcb.peercommunityin.org/articles/rec?id=150}
}
 \end{figure}

{\footnotesize\textbf{Note to the reader}: this article was originally conceived for readers who are mathematicians, with the objective of presenting an example of the application of mathematics in life sciences. It is for them that the first part has been designed, which readers already in the know can advantageously skip. On the other hand, the few mathematical formulas are likely to make the reading more difficult for people with a biology background, we hope that the explanations of their meanings will remain accessible to most people. The presentation of results in a field that is essentially multidisciplinary is always a little perilous.
}

~

We propose in this article a brief description of the work, over almost a decade, resulting from a collaboration between mathematicians and biologists from four different research laboratories, identifiable as the co-authors of the articles whose results are described here, and implicitely co-authors of this article, under the signature of Camille Noûs. This modeling work is part of population genetics, and is therefore essentially at the interface between mathematical tools, more particularly probabilistic ones, and biological data, more specifically genetic ones.

In a first part, we briefly present the theory of coalescence, which is the basis of our models, and the problems that this modeling tries to address. In a second part we describe our first results and the development of the IICR (Inverse of Instantaneous Coalescence Rate), the nodal point from which the different research paths we have been following are branching. The first results of these paths are summarized in the two following parts, one about the inference of demographic parameters of a structured population, the other about the consideration of selection.

\section{Coalescent theory in population genetics}

Population genetics is the study of the evolution of the genotypes in a population of living beings, under various evolutionary pressures such as mutation, selection or genetic drift. Initiated in the first half of the XX${}^{th}$ century with the work of the British statistician Ronald Fisher and the American geneticist Sewall Wright, it has seen the emergence of a backward model called \textbf{coalescent}, the first developments of which are due to the British mathematician John Kingman in the 80's (\cite{kingman1982coalescent}).

The coalescent theory consists in sampling individuals -- more precisely loci of individuals' genomes -- in the present population, and tracing their genealogies back in time, until a common ancestor, from two or more lineages, is found. The instants, backward to the past, when such common ancestors appear are called \textbf{coalescence times}, and are considered as random variables with values in $\N$ or in $\R^+$.

The mathematical object of interest is then the joint distribution of the various coalescence times of family trees, which allows to express the observable quantities in the genomes of a present population, as functions of this distribution. Those functions depend on genetic parameters (like mutation rate, recombination rate, selection rate) and demographic parameters (like sizes and numbers of sub-populations, migration rates between sub-populations). Note that in this paper, we will focus on loci separated by recombination events, and the topology of each tree of each locus won't matter. Observations of genetic sequences can then be used to infer genetic or demographic parameters using various statistical methods, and technological advances over the last two decades have made it possible to acquire huge masses of data, which can be used to refine existing models and develop new ones. 

\subsection*{Wright-Fisher model and Kingman coalescent}
\begin{figure}[!htb]
	\centering%
	\subfloat[ ]{\includegraphics[scale=0.21]{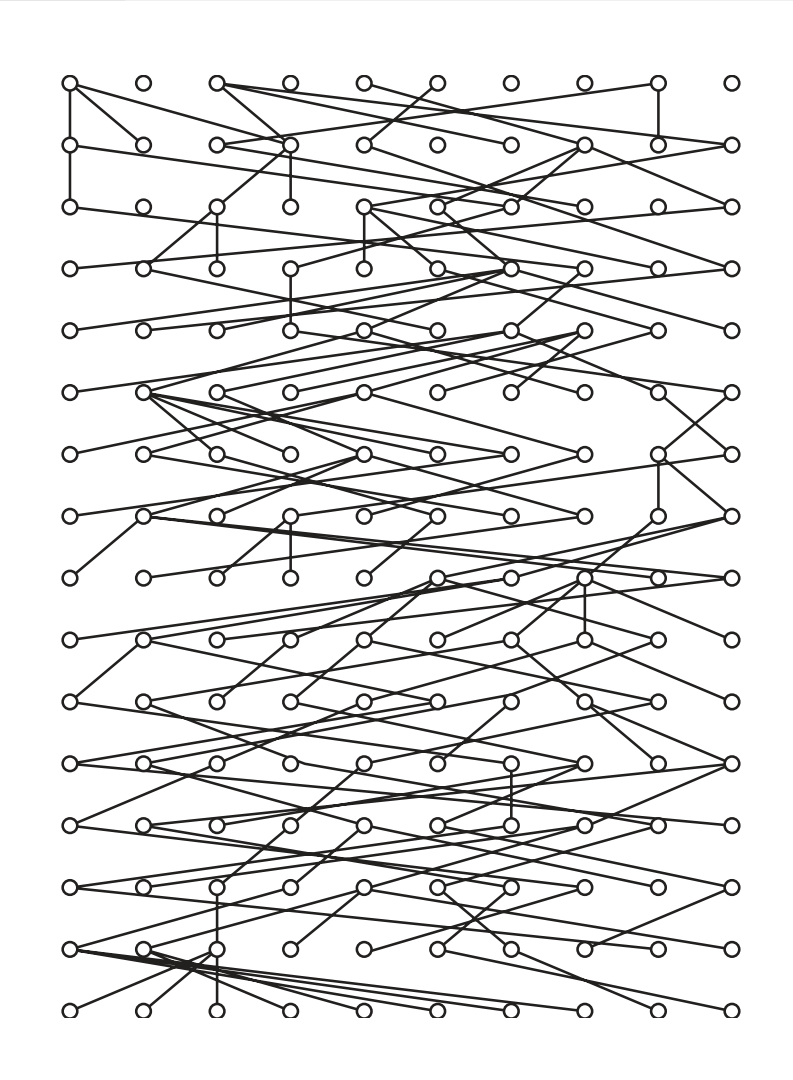}}~\quad~
	\subfloat[ ]{\includegraphics[scale=0.21]{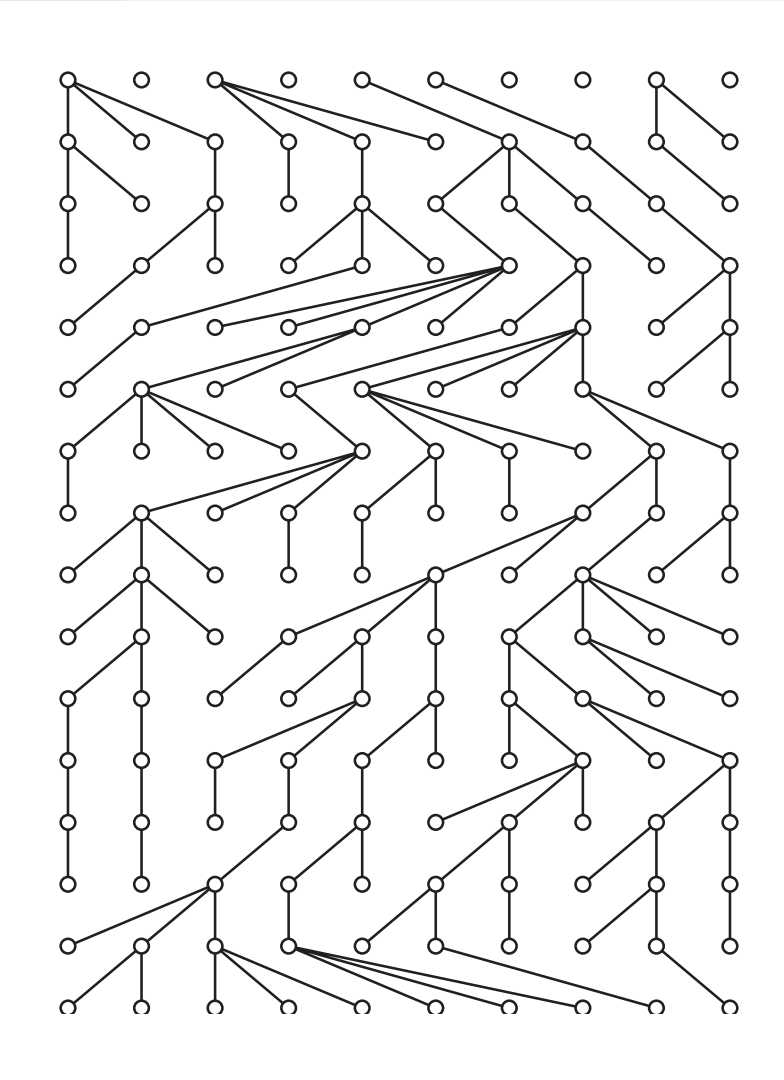}}~\quad~	
	\subfloat[ ]{\includegraphics[scale=0.206]{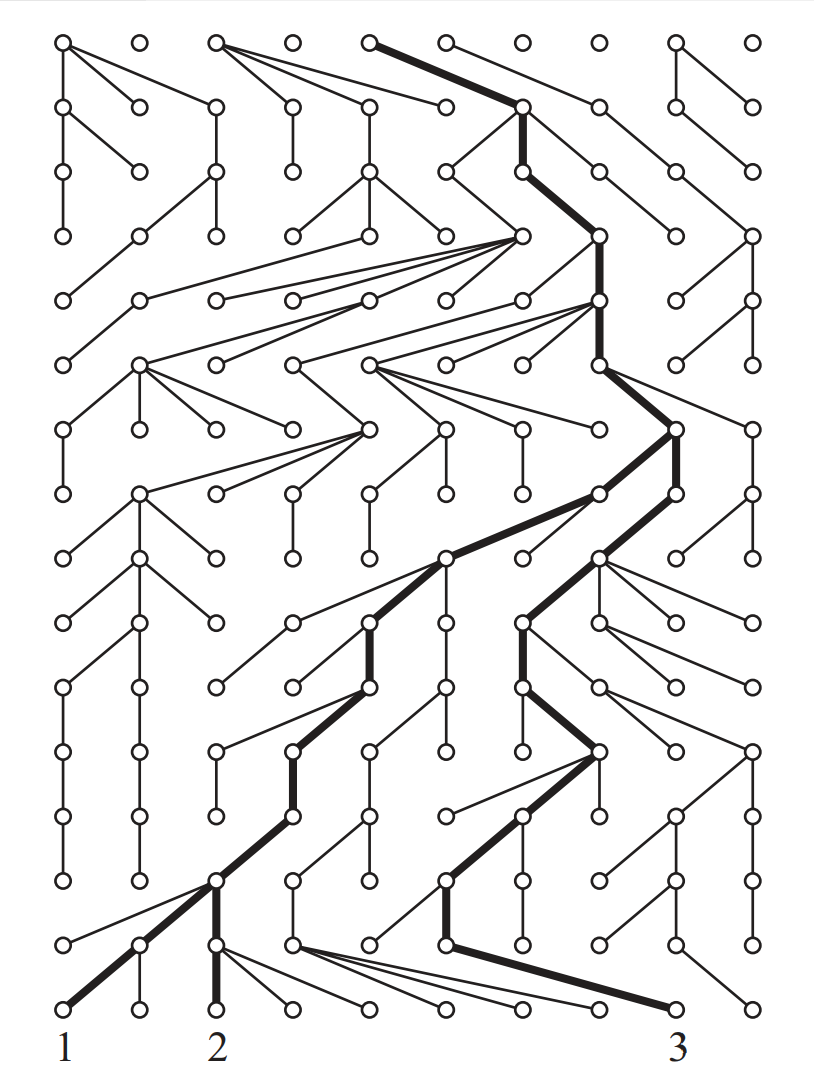}}
	\caption{\footnotesize{Here is a realization of the Wright-Fisher process on $16$ generations with a population size of $2N=10$. Panel \textbf{(a)} presents the evolution when each row is a generation, the individuals have been rearranged in panel \textbf{(b)} in order to highlight the family tree, and for panel \textbf{(c)} three individuals have been chosen in the last generation, their respective lineages having been put in bold. We see that the first coalescence between individuals $1$ and $2$ takes place two generations ago in the past, and that the last coalescence to the most recent common ancestor of individuals 1, 2 and 3 takes place nine generations ago in the past.}}\label{fig:WrightFisher}%
\end{figure}

More precisely, the Wright-Fisher model describes the evolution of a population of $2N$ individuals (the individuals can be genes or loci) with the following assumptions: in each generation, each individual independently generates a number of descendants following a Poisson distribution of the same constant parameter, with the offsprings completely replacing their parents, all conditioned by the fact that the size of the population must remain constant. The process can be described in an equivalent way backward in time: each individual of a given generation randomly chooses its parent in the previous generation in a uniform way. An illustration of the process is given in Figure \ref{fig:WrightFisher}\footnote{Figure from volume \cite{hein2004gene}.}.

If we now consider a pair of individuals in the last generation, and if we note $T_2$ the waiting time for the coalescence of the two lineages (going back in time), we have
$$
\P(T_2>i)=\left(1-\frac{1}{2N}\right)^i,
$$
and if we suppose $N$ to be large, by changing the time scale, we obtain the usual approximation of the geometric distribution by the exponential distribution
$$
\P(T_2> 2Nt)=\left(1-\frac{1}{2N}\right)^{2Nt}\sim \mathrm{e}^{-t}.
$$
We can thus generalize for the first renormalized coalescence time, which for convenience we keep noting $T_k$, even if it is not expressed in the same units, for $k$ individuals sampled in the last generation:
$$
\P(T_k>t)\sim \mathrm{e}^{-\binom{k}{2}t},
$$
and one thus obtains the coalescence tree, corollary of Kingman's coalescent, where the successive times of coalescence are independent, following exponential distributions of parameters equal to the binomial coefficients $\binom{k}{2}$.

\subsection*{Demographic complications of the coalescence model}
The coalescent defined this way is only valid in the context of a so-called panmictic population, i.e. without geographical structure (each individual randomly chooses its parent in the whole population), and with constant size. We will see how to generalize the coalescent in the case of a population of changing size, and in the case of a structured population.

\subsubsection*{Population size change}
If we consider that the size of the population can vary, by posing $N(i)$ the size of the population at generation $i$ in the past, and by considering the quantity 
\begin{equation}\label{lambda}
\lambda(t)=\frac{N(\lfloor 2Nt \rfloor)}{N(0)},
\end{equation}
 the relative size of the population in the past with the same temporal renormalization as in the previous section, it can be shown (see e.g. \cite{Tavare2004}, section 2.4) that under reasonable conditions of variation of $\lambda$ in the neighbourhood of infinity, the coalescence time $T_k$ of $k$ individuals sampled in the present satisfies
\begin{equation}\label{PTkch}
\P(T_k>t)=\exp\left( -\binom{k}{2}\int_{0}^{t} \frac{\mathrm{d}\tau}{\lambda(\tau)} \right).
\end{equation}
But unlike the panmictic case, the successive $T_k$ are no longer independent, which makes the global study of the tree more difficult.

\subsubsection*{Structured population}

To relax the assumption of panmixy opens the door to multiple ways of modeling population structuring. Classically, the global population is considered to be made up of subpopulations (called islands, or demes), each of which is panmictic, between which migration events may occur, with rates that may depend on each pair of islands. The demographic parameters of the model are thus: the number of islands $n$, the respective renormalized sizes $(s_i)_{i=1 \dots n}$ of the $n$ islands (again assumed constant in time), and the renormalized migration rates (to take into account the scaling already described which allows us to go to continuous time by assuming that populations are sufficiently large) $(M_{ij})_{i \neq j}$ between the islands $i$ and $j$.

The description of the coalescent tree thus becomes much more complex, but the information can be summarized, as Hilde Herbots-Wilkinson showed in 1994 in her landmark thesis work (\cite{Herbots1994}). If we note $\alpha=(\alpha_1, \dots, \alpha_n)$ the configuration where $\alpha_i$ represents the number of lineages present in the island $i$, then the coalescence process can be described by the infinitesimal generator $Q$ such that
\begin{equation}\label{Q}
Q(n_{\alpha},n_{\beta})=\left\{\begin{array}{cl}
\alpha_i\frac{M_{ij}}{2} & \text{if }\beta=\alpha-\epsilon^i+\epsilon^j \quad (i\neq j) \\
\frac{1}{s_i}\frac{\alpha_i(\alpha_i-1)}{2} & \text{if }\beta=\alpha-\epsilon^i\\
-\sum_i \left(\alpha_i\frac{M_i}{2}+\frac{1}{s_i}\frac{\alpha_i(\alpha_i-1)}{2}\right) & \text{if } \beta=\alpha \\
0 & \text{otherwise},\end{array}\right.
\end{equation}
where $\epsilon^i$ is the vector of size $n$ whose components are all zero except the $i$-th which is $1$. In order to have positive integers as indices of the matrix, we denote by $n_\alpha$ and $n_\beta$ the respective numbers of the $\alpha$ and $\beta$ configurations, once chosen a prior ordering of all possible configurations. The first line stands for a migration event, from island $i$ to island $j$, with the corresponding migration rate $\alpha_i\frac{M_{ij}}{2}$. The second means that two lineages in island $i$ coalesced, with rate $\frac{1}{s_i}\frac{\alpha_i(\alpha_i-1)}{2}$.

\subsection*{Genetic parameters, estimations and inferences}
From a genetic point of view, all these models are assumed to be neutral, i.e. not taking into account the possible influence of selection in the reproductive capacity of each individual. However, it is possible to easily incorporate the phenomena of mutation and recombination into these models, because they can be considered as events independent of the genealogical process. The classical assumptions are that each mutation or recombination event affects a different part of the genome (the so-called \textit{infinite site model}), and that the mutation and recombination rates are constant both in time and along the genetic sequences. 

\subsubsection*{Mutation and genetic diversity}
\label{sfs}
Mutation events are distributed on the genealogical tree according to a Poisson process, and we can link genetic diversity data, by observing for example the number of alleles of a given gene, or its distribution, and more generally the quantification of polymorphism, with the configuration of the tree (topology, length of branches) associated with the chosen model. By choosing a model, we can estimate the mutation parameter, and on the contrary, by assuming the mutation parameter to be known, we can estimate the lengths of the branches of the tree, and thus have information on the distributions of the coalescence times. Among the best known estimators, let us mention Watterson's $\theta_W$ based on the number of segregating sites (\cite{watterson1975number}) or the number of pairwise nuleotide differences (\cite{tajima1983evolutionary}).

\subsubsection*{Recombination and Sequential Markovian Coalescence}

The phenomenon of recombination is much more difficult to incorporate into these models than mutation, since it requires sexual reproduction, and at each recombination event the resulting genome is derived not from one parent but from two, thus exponentially increasing the number of ancestors involved for each lineage of individuals sampled in the present population. The ancestral recombination graph (ARG, see \cite{griffiths1996ancestral}) requires a computational treatment that is very quickly prohibitive when the sample size increases.

\begin{figure}[ht]
 \centering
 \includegraphics[width=0.8\textwidth]{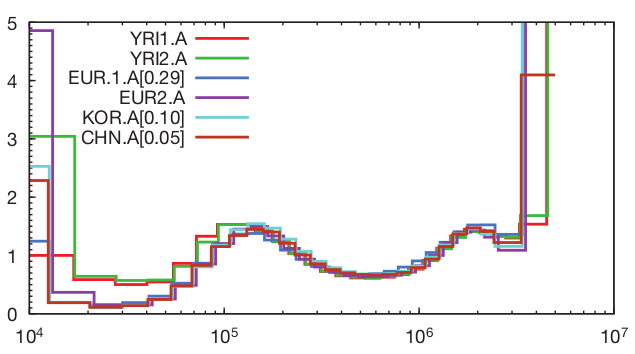}
 \caption{\footnotesize{Demographic inference obtained from human DNA from individuals of different populations (\cite{Li2011}). On the $x$-axis, the number of years in the past. On the $y$-axis, the renormalized size of the population, assumed to be panmictic.}}
 \label{fig:Estimation_PSMC}
\end{figure}

The work of Mc Vean and Cardin (\cite{mcvean2005approximating}) allowed, under an original hypothesis of the Markovian dependence property \textit{along the genome} (hence the so-called \textit{sequential} property), to greatly restrict the space to be explored for statistical inference methods.  Several demographic parameter inference software packages then emerged, including the famous PSMC (for \textit{Pairwise Sequentially Markovian Coalescent}, \cite{Li2011}), which has been widely used since 2011, and allows to estimate the variation of the population size (noted $\lambda(t)$ in equation \eqref{lambda}), using the genetic data from a diploid individual only (fully sequenced genome), see for example Figure \ref{fig:Estimation_PSMC}.

\section{Consideration of population structure, central role of the IICR}

Some work in the first decade of this century (\cite{wakeley2001coalescent}, \cite{chikhi2001estimation}, \cite{chikhi2010confounding}) highlighted the effect that a structured population could generate on statistics assuming panmixia, with notably the detection of false bottleneck signals in some cases. This can be problematic, for instance for conservation population issues. It is not always easy to quantify some kind of degree of structuration, there exists a large bibliography on this topic (see for instance \cite{chakraborty1993analysis} or \cite{excoffier2004analysis}).

After a preliminary study which consisted in analyzing a simple case of comparison of an island model and a size change model (\cite{mazet2015demographic}), we obtained a first result, which has become a nodal point of our subsequent research and which we present here.

\subsubsection*{The IICR and the change in size}
We have highlighted in \cite{Mazet2016} the following result. Considering that whatever the chosen demographic model is, the coalescence time $T_2$ of the lineages of two individuals chosen in the present population is a random variable with values in $\R^+$. This variable thus can be considered as a lifetime of density $f_{T_2}$, and as such, admits a ``failure rate''  which here translates into an \textbf{(instantaneous) rate of coalescence} equal to

$$
\mu(t)=\frac{f_{T_2}(t)}{\P(T_2>t)}.
$$
The density of $T_2$ can thus always be written 
$$
f_{T_2}(t)=\mu(t)\exp\left( -\int_0^t \mu(\tau) \mathrm{d}\tau \right),
$$
hence
\begin{equation}\label{PT2gen}
\P(T_2>t)=\exp\left( -\int_0^t \mu(\tau) \mathrm{d}\tau \right).
\end{equation}

\begin{figure}[ht]
 \centering
 \includegraphics[height=8cm]{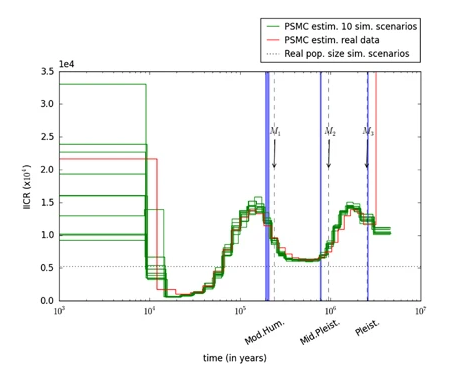}
 \caption{\footnotesize{In red the PSMC of real data from a human (CHN.A in Figure \ref{fig:Estimation_PSMC}). In green the PSMC of 10 simulations of the same model in islands, of constant size, with three changes of migration rate represented by the vertical dotted lines, at 2.52 Myr ago, 0.95 Myr ago and 0.24 Myr ago. The blue shaded areas correspond to the beginning of the Pleistocene at 2.57–2.60 Myr ago, the beginning of the Middle Pleistocene at 0.77–0.79 Myr ago and the oldest known fossils of anatomically modern humans at 195–198 kyr ago. Following \cite{Li2011}, we assumed that the mutation rate was $\mu=2.5 \times 10^{-8}$ and that generation time was 25 years. We also kept their ratio between mutation and recombination rates. Each deme had a size of 530 diploids and the total number of haploid genomes was thus constant and equal to 10 600.} }

 \label{fig:hum_niles}
\end{figure}

If we now bring equation \eqref{PT2gen} together with the particular case $k=2$ of equation \eqref{PTkch}, we realize that in the panmictic case, the change in size $\lambda(t)$ is exactly equal to $\frac{1}{\mu(t)}$, which is thus the inverse of the instantaneous coalescence rate, noted by the acronym \textbf{IICR}. Two important consequences can be drawn from this observation:
\begin{enumerate}
\item The sole data of the $T_2$ distribution cannot be informative of the demographic model, if we don't know if it is structured or not, since whatever it is, there is always a panmictic model which will provide exactly this $T_2$ distribution. Indeed, it is sufficient to choose the inverse of the coalescence rate as the size change. 
\item What software like PSMC infers, on data from a single diploid genome and however long it may be, is the IICR associated with the demographic model, which is usually \textbf{not the change in size} of the population when it is structured.
\end{enumerate}
Taking the second consequence further, as a proof of concept we built a constant size demographic model based solely on a symmetric island model, with the number of islands also constant, where only the migration parameter is allowed to vary. As we can see in Figure \ref{fig:hum_niles} extracted from \cite{Mazet2016}, the PSMC output on data simulated under this model is very similar to that on real human data.

There is obviously no question of claiming that the human population is structured in symmetrical islands and that its population size has remained constant over the course of evolution, but this example prompts us to question the interpretation of the IICR, which is the object inferred by the PSMC, and shows that it is necessary to investigate further, before drawing any conclusions about the demographic history of a population.

\subsubsection*{Links with previous results}

Beyond the formal framework, which confirms what  simulations in \cite{wakeley2001coalescent}, \cite{chikhi2001estimation} or \cite{Chikhi2010} had already shown, i.e. bottleneck or population expansion signals depending on whether the lineages are sampled in the same demes or not (see also \cite{peter2010distinguishing} or \cite{heller2013confounding}), the IICR also makes it possible to establish links with works on various concepts of effective sizes of a structured population.

Indeed, the IICR of a stationary structured model, after a transitory phase which depends on the sampling, always converges towards a plateau which could be considered as an "asymptotic" effective size and whose value is calculable in an exact way, according to the demographic parameters (number and size of the demes, rate of migration between the demes). It turns out that for the n-island and for fairly large migration values, in the first approximation this value is equal to the metapopulation size $Nn$ ($n$ being the number of subpopulations and $N$ the size of each of them), and in the second approximation (by adding the next term of the Taylor expansion) this value is equal to $N(n+\frac{(n-1)^2}{nM})$, which is the effective size computed in \cite{nei1993effective} for this model.

Let us also note in the work of \cite{whitlock1997effective} a certain similarity of approach, but for the case where the unit of time remains the generation, and where the \textit{eigenvalue effective population size} defined by \cite{ewens1982concept} is studied in connection with "their" Inverse of Instantaneous Coalescence Rate.

\subsubsection*{Influence of the sample size}
\label{Tk}
From the first consequence drawn above, we explored theoretically what the data of a third lineage could bring as additional information. We then showed that in the simple case of a population structured in islands, then adding the information of the $T_3$ distribution to the $T_2$ distribution is enough to distinguish this model from the panmictic model having the same $T_2$ distribution, thus the same IICR (\cite{Grusea2018}).

This result provides theoretical evidence that a sample size strictly greater than two is sufficient to distinguish an island structured model from a panmictic model, but initial attempts to move into practice have not yet been successful, because of the precision required, which often blends into the noise of the real data. 

\subsubsection*{Structure and IICR: sampling strategy}

\begin{figure}[ht]
 \centering
 \includegraphics[width=0.6\textwidth]{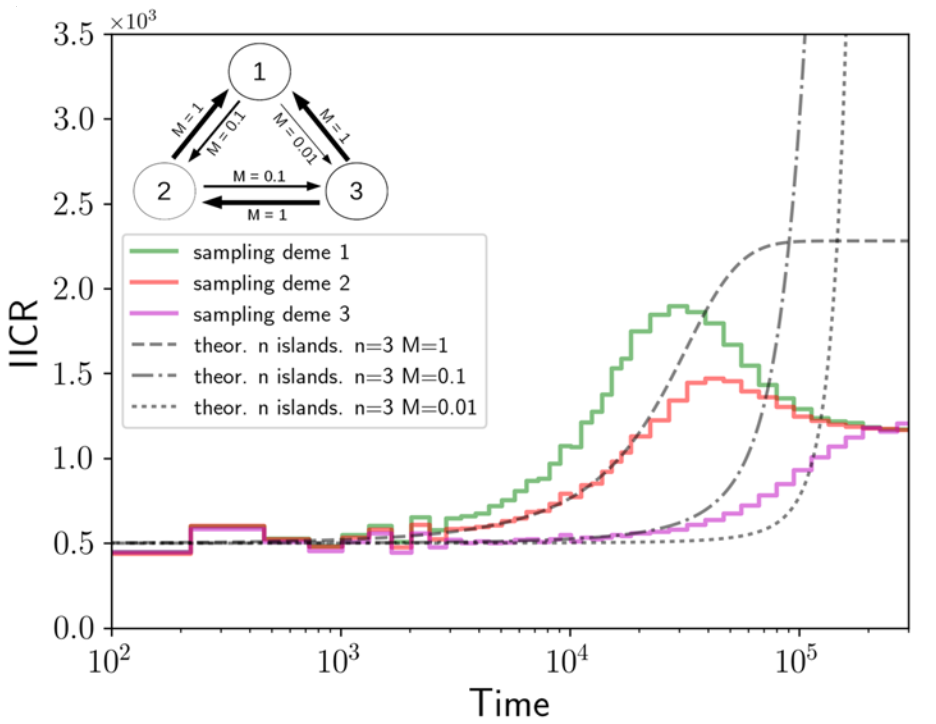}
 \caption{\footnotesize{The size of each island is constant. However, the IICRs of each pair of sequences are time-varying functions, and these functions even may not be monotonic. Furthermore, they differ depending on the island from which the pair is sampled. This Figure comes from \cite{Chikhi2018}.}}
 \label{fig:3iles}
\end{figure}

Exploratory work was then done (\cite{Chikhi2018}), using simulated data, to find out what signatures are left on the IICR produced by different types of structured models, and thus indirectly (or directly when dealing with software of the same type as PSMC) on the false signals of size changes that these models generate. As an illustration, we present in Figure \ref{fig:3iles} extracted from this paper, the simulated IICRs in a model with three islands and asymmetric migration rates, by sampling a diploid individual in each of the islands. We see that not only the structure of the model can give false signals of size change for software assuming panmixia, but also the IICR is \textbf{dependent on the sampling location}, for the same demographic history. Indeed, sampling a diploid individual in island 3 would provide a signal of monotonic population decrease, whereas sampling it in island 1 or 2 would provide a signal of first expansion, followed by reduction, with different magnitudes. This finding also deserves to be further explored for use in model selection.

\subsubsection*{The IICR as a model validation}
\begin{figure}[!h]
 \centering
 \includegraphics[width=0.6\textwidth]{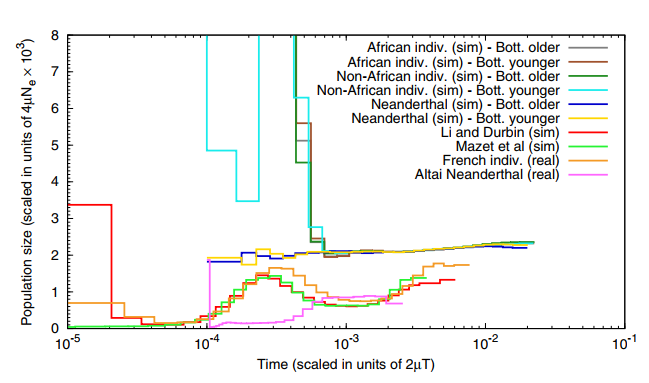} 
 \caption{\footnotesize{The PSMCs of real modern human and Neanderthal data (the last two in the order of the legend), set against the PSMCs of simulated data with parameter values used by the authors of different proposed models in \cite{yang2012ancient}~: the African, Non-African and Neanderthal individuals were simulated under the model of recent admixture with a bottleneck that was either older or younger than the admixture event. And for the record, the panmictic model suggested by \cite{Li2011} and the fictional structured model proposed by \cite{Mazet2016}. For more detailed explanation, see \cite{Chikhi2018}.}}
 \label{fig:lounes}
\end{figure}

The IICR can also be used as a \textbf{summary statistic} of a given model, for validation or rejection. In the same paper \cite{Chikhi2018} we thus tested a number of models proposed in the literature for human evolution, for \textit{Homo sapiens} as well as for \textit{Homo neandertalensis}. Simulating the IICR of some of these models allowed us to discard them, as the IICR produced differed radically from that estimated by PSMC on human data. For example the Figure \ref{fig:lounes} illustrates this situation for the models proposed in \cite{yang2012ancient}, which was supposed to prove that admixture between humans and Neanderthals explained better the data than ancient structure.

\subsubsection*{IICR and structured coalescent}

On the theoretical side, work has been undertaken to calculate, as precisely as we like, the IICR of any structured model (\cite{rodriguez2018iicr}). The modeling initiated by Herbots provides a set of infinitesimal generators of Markov processes (see formula \eqref{Q}), and it is possible to exploit the semi-group property of the exponentials of these matrices. Indeed, changes in some parameters of the structured models, such as island sizes or migration rates, leave the state space of the process unchanged, so the matrices can be piecewise constant functions of time. For example, if we suppose that at a date $T$ in the past some of the parameters $M_{ij}$ or $s_i$ change, and if we note by $Q_0$ the generator for the time $0\leq t\leq T$ and by $Q_1$ the one corresponding to the time $t>T$, the transition semi-group of the Markov chain can be written as follows:
\[
P_{t} =
\begin{cases}
  \mathrm{e}^{tQ_{0}}, & \text{ if } t\leq T \\
  \mathrm{e}^{TQ_{0}}\mathrm{e}^{(t-T)Q_{1}},\ \ & \text{ otherwise}.
\end{cases}
\]
In particular, the distribution of $T_2^{\alpha}$, coalescence time of two lineages starting from a $\alpha$ configuration, is deduced from
\[
\mathbb{P}(T_{k}^\alpha \leq t)=P_t(n_{\alpha}, n_c),
\]
where $n_\alpha$ is the number of the state corresponding to $\alpha$, and $n_c$ the number of the coalescence state. Its density is then equal to $f_{T_{k}^\alpha}(t) = P'_{t} (n_{\alpha}, n_c),$ where
\[
P'_{t}  =
\begin{cases}
  \mathrm{e}^{tQ_{0}}Q_{0}, & \text{ if } t< T \\
  \mathrm{e}^{TQ_{0}}\mathrm{e}^{(t-T)Q_1}Q_1,\ \ & \text{ otherwise}.
\end{cases}
\]

\begin{figure}[ht]
 \centering
 \includegraphics[height=6cm]{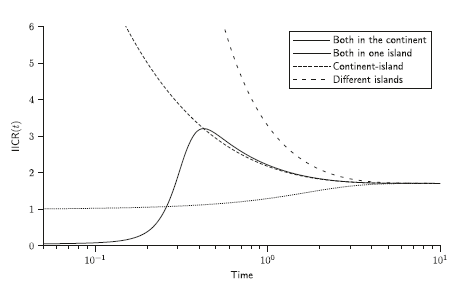}
 \caption{\footnotesize{Theoretical IICR of the structured model with a continent of size $1$, three islands of sizes $\frac{1}{20}$, and migration rates proportional to sizes between islands and the continent (no migration between islands). We find, as in the simulations in \cite{Chikhi2018}, the importance of sampling location, as well as the obvious false signals of population size changes that software such as PSMC could infer, here the demographic model being constant over time.}}
 \label{fig:ilescont}
\end{figure}

These explanations allow to numerically determine the theoretical IICRs of a large number of structured models, such as the continent-islands model in Figure \ref{fig:ilescont}, with possible changes in demographic parameters, such as subpopulation sizes or migration rates.

\begin{figure}[ht]
 \centering
 \includegraphics[width=0.8\textwidth]{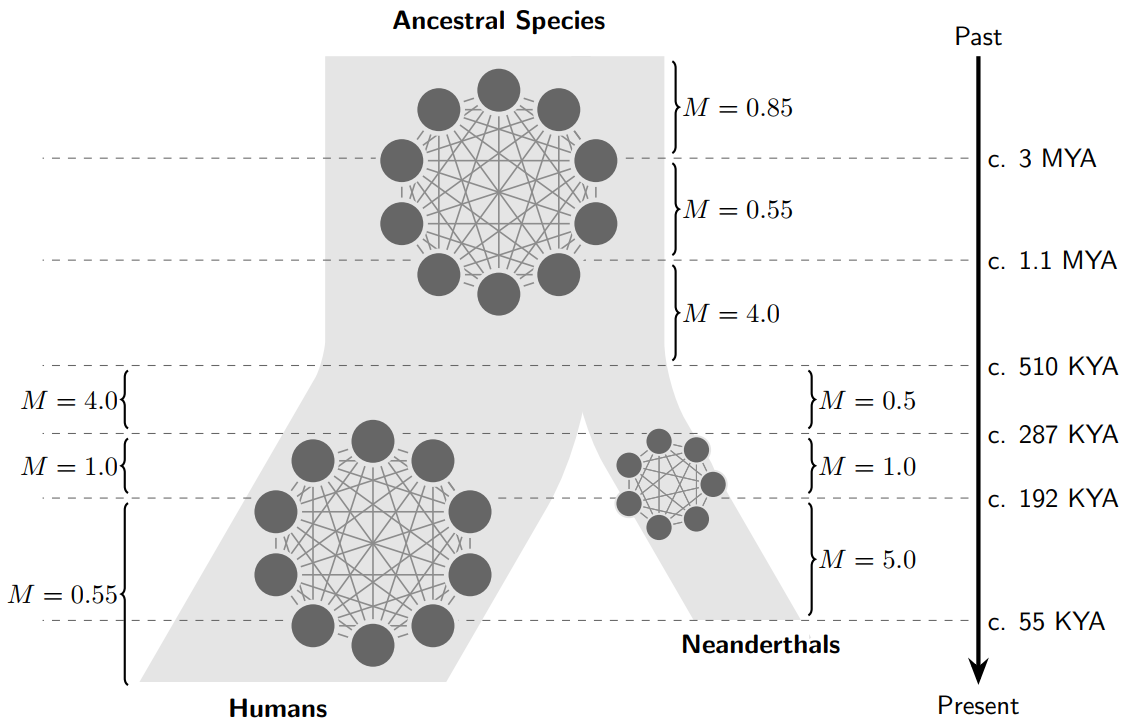}
 \caption{\footnotesize{Hypothetical scenario presenting humans and Neanderthals as structured species derived from an unknown Homo species that was itself structured, and without any gene flow between the two species after the split. The times at which gene flow (M) changed are indicated by horizontal lines.}}
\label{mod_sapiens_neand}
\end{figure}

\begin{figure}[ht]
 \centering
 \includegraphics[width=0.8\textwidth]{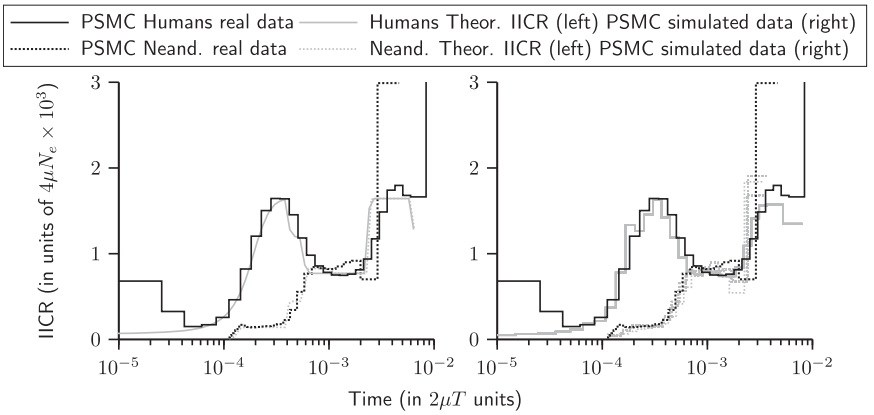}
 \caption{\footnotesize{Superposition of the PSMCs of real Neanderthal and Sapiens data, with the theoretical IICRs of the proposed structured model (left) and the PSMCs of the simulated data from this structured model (right).}}
\label{fig:sapiens_neand}
\end{figure}

Also as a proof of concept, an extended fictional model of human evolution was proposed, integrating Neanderthals alongside modern humans in the same constant size structured model, with only the migration coefficients allowed to change. The model is described in Figure \ref{mod_sapiens_neand}, and the simulated PSMCs are presented in Figure \ref{fig:sapiens_neand} with PSMCs of real data. We then see that relatively simple structured model can explains the PSMCs without change of the metapopulation sizes before and after the split between Neanderthals and modern humans from an unknows \textit{Homo} species.

\section{Inference of parameters in a structured model}
\subsection*{IICR of $T_2$ for a structured model}
The theoretical possibility (presented in \cite{rodriguez2018iicr}) of numerically computing the IICR of two sampled lineages in any structured model, as a function of the demographic parameters such as the number of islands, the successive island sizes, the successive migration rates, and the times of change of parameters like sizes or migration rates, opens the way to estimate these parameters from IICRs inferred from real data, e.g. via the inevitable PSMC. The challenges of complexity and computation time, even in the simplest case of the symmetric island model, have been overcome thanks to the work of Armando Arredondo, part of his PhD thesis (\cite{ArredondoThesis}), with the design and realization of a software for inferring such parameters, called SNIF (Structured Non-stationary Inferential Framework), presented in \cite{arredondo2021inferring}.

\begin{figure}[!ht]\centering%
\includegraphics{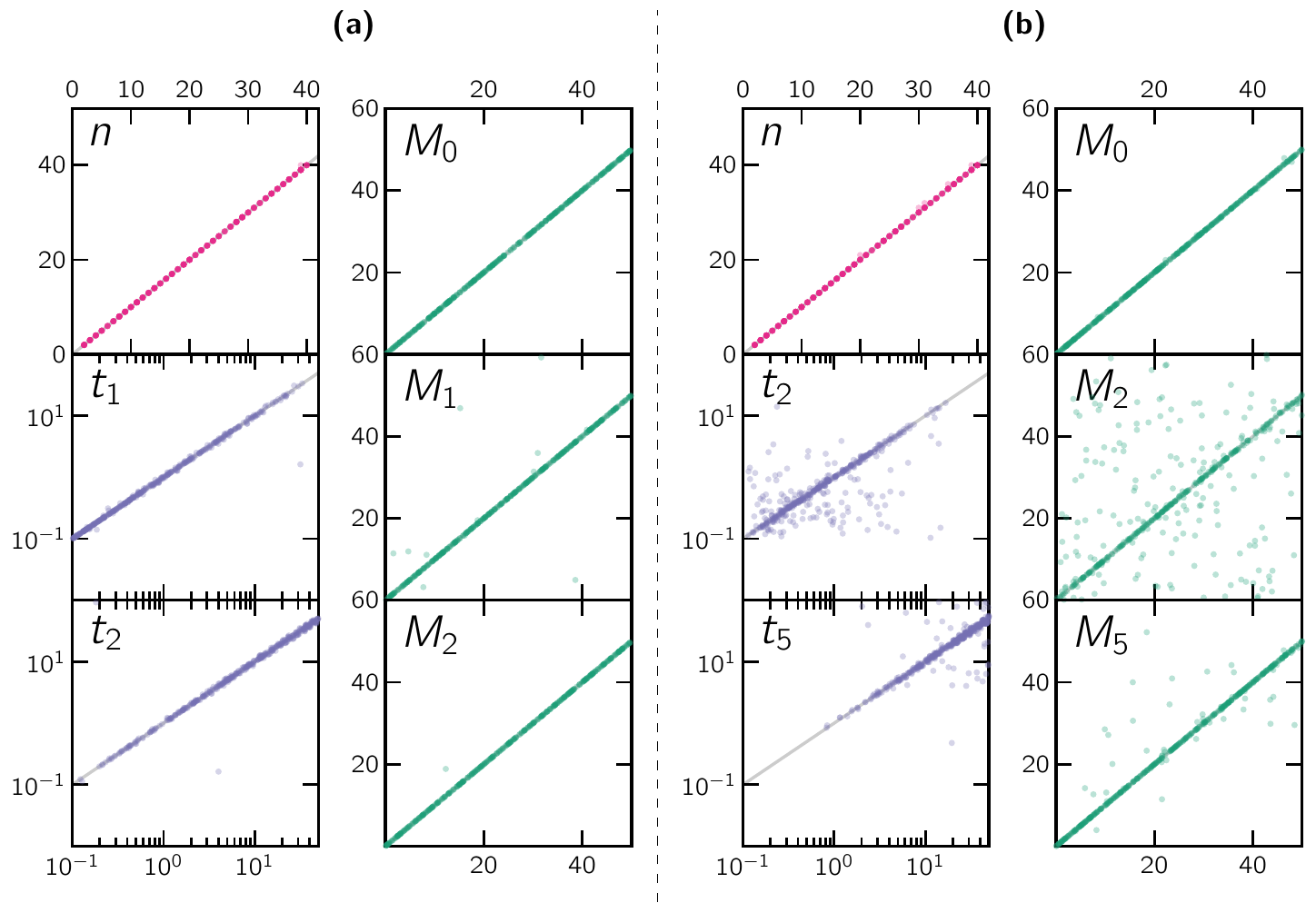}%
\caption{\footnotesize{Scatter plots of simulated and inferred parameters. $n$ is the number of islands, $t_i$ the $i$-th change of value of the migration parameter, and $M_i$ the $i$-th value of the latter. Panel \textbf{(a)} corresponds to scenarios with $c=3$~components, and \textbf{(b)} to scenarios with $c=6$~components. The different sub-panels represent the simulated (horizontal axis) versus inferred (vertical axis) parameter values for all the parameters (or a representative selection of parameters in the case of panel \textbf{(b)}) of $L=400$ unscaled simulated scenarios.}}\label{fig:SNIF}%
\end{figure}

Testing this software on simulated data revealed a first problem of identifiability between subpopulation sizes and migration rates. Second, if we want to obtain an acceptable level of precision for the estimation of the migration rates, the number of different values over time (this number is called the number of ``components'') should not be too large, generally not more than 5 or 6. On the other hand, the estimated number of subpopulations is extremely reliable. A synthesis of these first results can be seen on Figure \ref{fig:SNIF}.

\begin{figure}[!ht]\centering%
\includegraphics[width=0.95\textwidth]{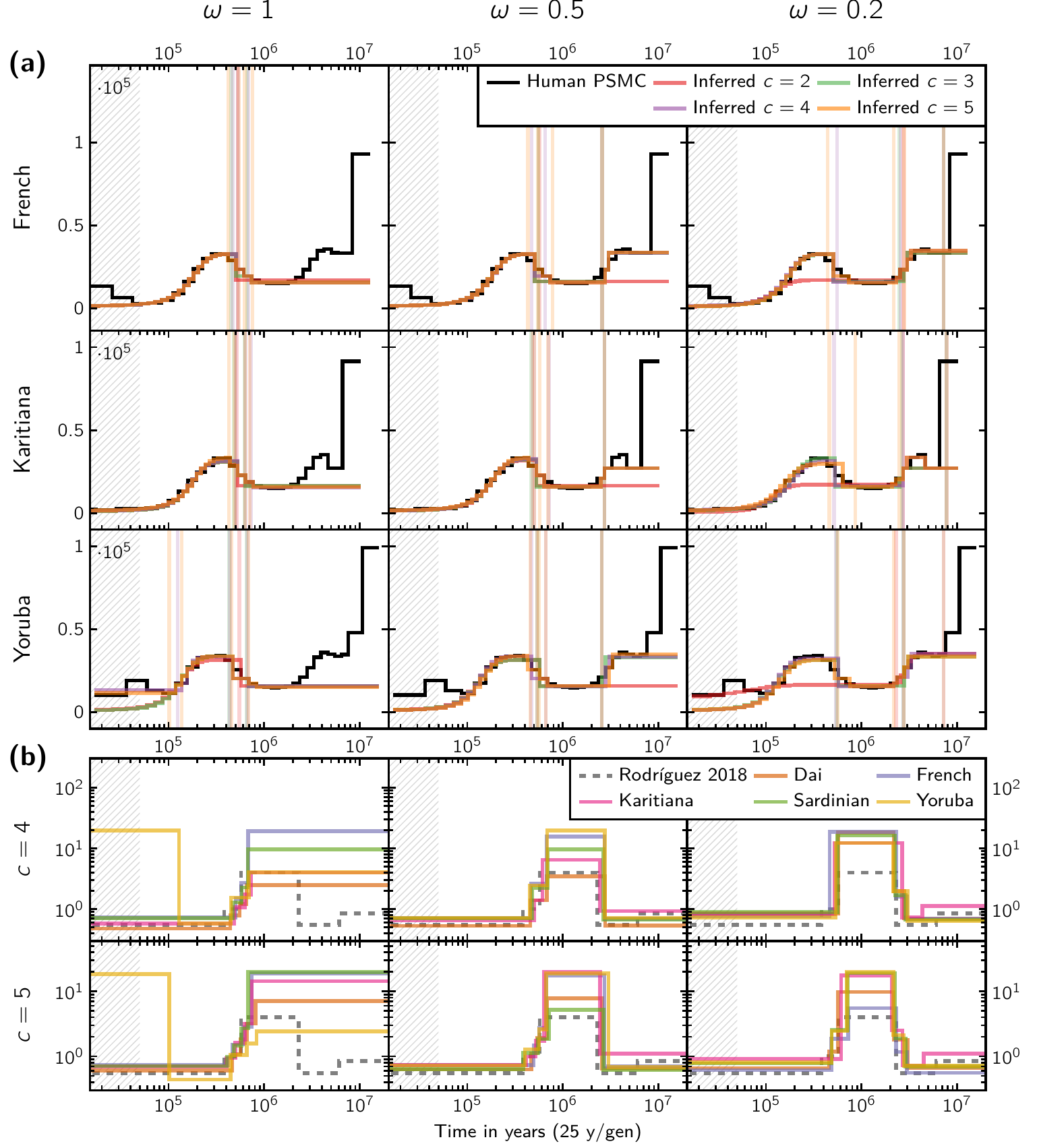}%
\caption{\footnotesize{Results of performing demographic inference on three representative human PSMC curves. Panel \textbf{(a)} shows the various IICR plots inferred for the different populations, numbers of components $c$ and weight parameters $\omega$ used, together with the target IICR curves (or PSMC plots) on which these estimations are based. Panel \textbf{(b)} shows the connectivity graphs for the same set of inferred scenario. As a reference point, the connectivity graph of the scenario proposed in \cite{rodriguez2018iicr} is also shown. The vertical axes represent migration rates ($M$).}}\label{fig:SNIF_PSMC}%
\end{figure}

An application on human data also allows to find a good model in symmetrical islands which explains surprisingly well the graph produced by the PSMC, see Figure \ref{fig:SNIF_PSMC}.

This method has already been used to contribute to the study of the evolution of species of microcebes, Malagasy lemuriform primates (\textit{Microcebus murinus} and \textit{Microcebus ravelobensis}). The results are published in (\cite{teixeira2021}). Other data on other species of mammals are being analysed using this software.

\subsection*{Increase of the sample size}
A natural way to increase the precision of the estimation of demographic parameters, is natural to increase the size of the statistical sample. For instance, the distribution of alleles frequency over the gene, generally called SFS (for \textit{Site Frequency Spectrum}), is commonly used in population genetics, since it's easy to extract values from real data. Nevertheless, the average SFS is theoretically known only for a panmictic model (see for example \cite{griffiths1998age}), but the calculation becomes of great combinatorial complexity for any structured model. In the case of the island model, Armando Arredondo has just completed the theoretical treatment of obtaining the average SFS for any value of $k$, as well as the feasibility in computation time for a sample size of $k \leq 26$ in the current state of computing capabilities (\cite{ArredondoThesis}, chapter 3). It now remains to implement this algorithm in the inference software.

\subsubsection*{IICR$_k$: derivation and first results}
Coming back to the coalescences times $T_k$ of such a sample of size $k$, their IICR (noted here IICR$_k$) for $k>2$ is theoretically easily computable thanks to the infinitesimal generator of equation \eqref{Q} and the extensions exposed in section \ref{Tk}. Indeed, the IICR$_k$ of the first time $T_k$ of coalescence of $k$ lineages can be defined in the same way as the IICR (which is in fact the IICR$_2$):
$$
\text{IICR}_k(t)=\frac{f_{T_k}(t)}{\P(T_k>t)}.
$$
While we know that in the panmictic case we have 
$$
\forall k\geq 2, \forall t>0, \qquad \text{IICR}_k(t)=\frac{1}{\binom{k}{2}}\text{IICR}_2(t),
$$
this is not the case for a structured model (as we formally showed for the symmetric island model in \cite{Grusea2018}). There already exist powerful methods to estimate the IICR$_k$ of real genomic data of sample size $k$, notably the extensions of the PSMC, called MSMC (for \textit{Multiple Sequentially Markov Coalescent}, see \cite{Schiffels2013}). The practical problem comes from the fact that the larger the sample size, the shorter the coalescence time, and thus the fewer the genomic traces on the data, because the number of mutation and recombination events decreases very quickly, and falls below the acceptable threshold for the statistical estimation to be satisfactory.

Still, for small values of $k$, (let's say from $k=3$ to $k=8$), some results are being obtained in a work in progress, which show that IICR$_k$ could provide an efficient tool to distinguish a structured population from a panmictic one.

\section{IICR and consideration of selection}
All the models we have discussed so far are so-called neutral models, i.e. they do not take into account the selection pressure that individuals undergo at some loci to increase the frequency of new advantageous alleles (positive selection), decrease the frequency of new deleterious alleles (negative selection) or maintain polymorphism (balancing selection). A large bibliography deals with the detection of positive selection events (the so-called selective sweeps see  
the review in \cite{walsh2018evolution}), and many studies have investigated the influence of demography on sweep detection (including for instance \cite{jensen2005distinguishing} or \cite{bonhomme2010detecting}, but there are many others). On an other side, fewer and more recent papers studied the impact of selection on demography inference, like \cite{ewing2016consequences} or \cite{johri2020impact} for negative selection, and for sweeps, \cite{schrider2016effects} or \cite{harris2020genomic}.

One classical way to model the long-term combined effect of selection and genetic linkage (i.e. linked selection) on genomic sequences is to assume that some portions of the genome undergo a recurrent higher impact of selection due to their local gene content or recombination rate, resulting in an effective size that differs from that in neutral areas (see \cite{hill1966effect}, \cite{charlesworth2009effective}, \cite{gossmann2011quantifying} or \cite{jimenez2016heterogeneity} for example). Under this approach, regions under positive or negative selection are modelled by a lower effective size, while regions under balancing selection are modelled by a higher. This is a way to make the coalescence rate variable over the genome, this rate being linked to the reproductive capacity. 

\begin{figure}[ht]
 \centering
 \includegraphics[width=\textwidth]{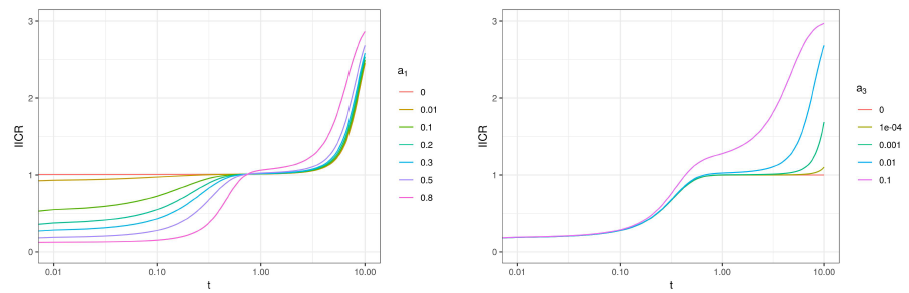}
 \caption{\footnotesize{Example of IICR for $K=3$, $\lambda_1=0.1$, $\lambda_2=1$ and $\lambda_3=3$. On the left we set $a_3=0.01$ and on the right $a_1=0.5$. The value $\lambda_3$ determines the limit, and $a_3$ the speed of convergence, with more or less pronounced transient plateaus depending on the other values.}}
 \label{fig:IICR_sel_pan}
\end{figure}

Since the IICR is directly related to the coalescence rate, it is natural to explore the influence of modeling selection by the variability of the effective size along the genome on the IICR (\cite{boitard2022heterogeneity}). A theoretical calculation allows us to show, under the simple hypothesis of a panmictic population, that if we assume the existence of $K$ classes of the genome under respective effective sizes equal to $\lambda_i=\frac{1}{\mu_i}$ for $i=1\dots K$, each corresponding to a proportion $a_i$ of the genome (with $\sum_i^K a_i=1$) then the IICR is
$$
\text{IICR}(t)=\frac{\sum\limits_{i=1}^K a_i\mathrm{e}^{-\mu_i t}}{\sum\limits_{i=1}^K a_i\mu_i\mathrm{e}^{-\mu_i t}},
$$
and a basic calculation indicates that for all values of $K$, $a_i$ and $\lambda_i$, this IICR is always increasing on $\R^+$, with 
$$
\text{IICR}(0)=\frac{1}{\sum\limits_{i=1}^K a_i\mu_i} \quad \text{and} \quad \lim_{t \to +\infty}\text{IICR}(t)=\max_{i=1 \dots K}\lambda_i.
$$
It is thus mathematically shown that the IICR will exhibit a decline signal of population size, as it can be seen in Figure \ref{fig:IICR_sel_pan}. An interpretation could be that in recent past the coalescent rate is high (and consequently the effective size appears to be small) because of coalescent events happening in zones of small effective size, and in distant past are remaining the coalescent events occuring in zones of large effective size. Figure \ref{fig:IICR_sel_pan} shows also that under these assumptions, the largest effective size present in the genome, even under a very small proportion, has a significant influence on the growth of the IICR as a function of time from the present to the past. The magnitude of this variation may be surprising, but the simulations also show, especially in the presence of structure in the population, that it may be statistically difficult to detect the final plateau, given the small number of corresponding coalescence events.

\begin{figure}[ht]
 \centering
 \includegraphics[width=\textwidth]{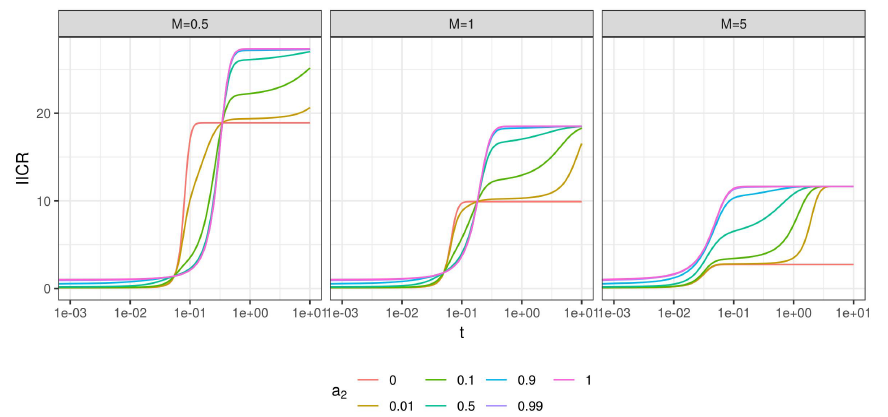}
 \caption{\footnotesize{Example of IICR for a symmetric 10-island model, migration rate $M$, and a genome with $K=2$ effective size classes $\lambda_1=0.1$ and $\lambda_2=1$ of relative proportions $a_1$ and $a_2$.}}
 \label{fig:IICR_sel_str}
\end{figure}

It is possible to generalize the derivation of the analytic expression of the IICR if the population is structured. If we denote by $f_i(t)$ and by $a_i$ respectively the density of the coalescence time $T^i_2$ and the proportion of the $i$-th of the $K$ classes of the genome, let's recall that in the panmictic case we have $\P(T^i_2>t)=e^{-\mu_i t}$ and $f_i(t)=\mu_ie^{-\mu_i t}$, so if the population is not panmictic, the formula becomes
$$
\text{IICR}(t)=\frac{\sum_{i=1}^K a_i \P(T^i_2>t)}{\sum_{i=1}^K a_i f_i (t)},
$$
and thanks to our previous work on island-structured models, we can combine the effects of structure and of selection and numerically calculate the corresponding IICRs (see Figure \ref{fig:IICR_sel_str}). We can then see that, even if we find the same monotonic pattern as in the panmictic case, the structure hides partly the growth towards the limit value, which is reached quite far in the past.

Throughout this section, the coalescence rate was considered variable along the genome, but was assumed to be constant over time to model the long-term average effect of selection. The IICR mathematical framework also allows to model \textit{transient selection}, which was done in \cite{boitard2022heterogeneity} for one specific scenario, again from the point of view of the impact on the PSMC. In order not to overload this section, we let the interested reader refer to this article.

Finally, it should be noted that this study only uses information from the $T_2$ distribution, which may explain why our results contradict some of the literature by showing signals of recent population decline. Indeed, for larger sample sizes, especially when using the SFS, negative selection leads to an excess of singletons, which under the neutral hypothesis leads to a signature of a recent population expansion.

\section{Conclusion and prospects}

In summary, the IICR of $T_2$, despite its intrinsic limitations, on the one hand because it is a distribution of a variable which is not directly observable, and on the other hand because it is based on a sample of size 2, proves to be an extremely fertile object of modeling. The matrix writing of this time function\footnote{It should be noted that the mathematical objects studied in our work are part of a general formalism coming from the theory of phase-type distributions, see for example \cite{hobolth2019phase}.} facilitates precise numerical calculations (\cite{rodriguez2018iicr}), and makes accessible powerful inference methods, such as the recently developed SNIF program (\cite{arredondo2021inferring}).

The IICR also sheds new light on a concept with which it is naturally associated, that of \textit{effective size}, which is the source of an abundant literature (see, for example, \cite{charlesworth2009effective}, or the recent \cite{waples2022n} for a complete review), and is subject to many different, sometimes contradictory, interpretations. The starting point is the evidence of a direct correspondence between the IICR and the population size in panmictic condition. But a first level of hypothesis complexification, with the introduction of a population structuring, quickly leads to erroneous conclusions about size changes. We can thus observe variations in the effective size that do not correspond to those of the real size, or even that are in the opposite sense (\cite{rodriguez2018iicr}, \cite{Chikhi2018}). 
It is nevertheless difficult to quantify a level of structuration from which the problem highlighted could be qualified as serious. Moreover, it seems obvious that if structure is omnipresent, so is real variation in population size, and the combination of the two phenomena, as well as their variability over time, makes the inference of demographic parameters and the interpretation of results more complex.
Similarly, it is useful to know how to detect possible effective size variations, induced by the introduction of genomic areas under selection, positive or negative (\cite{boitard2022heterogeneity}). Another crucial contribution of the IICR is to highlight the importance of the sampling strategy (\cite{Chikhi2018}), the exploitation of which should allow new methods for model selection.

Among the other more immediate prospects, we note the continuation of the theoretical study of the IICR for models a little more elaborate than the symmetric island model (first of all the asymmetric island model, the island-continents model, or even the one or two dimensional stepping-stone model), with the objective of highlighting the influence of the values of the demographic parameters on the variations of the IICR, via the eigenvalues of the associated infinitesimal generator. Finally, in order to increase the predictive and explanatory capacities of our models by enlarging the sample size, we have to develop the inference methods from the existing one (\cite{arredondo2021inferring}), by incorporating on the one hand the average SFS of an island model, and on the other hand the IICR of $T_k$.

\small{
\subsection*{Acknowledgements}

As we said in the preamble, the results presented here are the result of a direct collaboration of many researchers in biology and mathematics, a list of which can hardly be exhaustive and can be found here : Armando Arredondo Soto, Simon Boitard, Alexandre Changenet, Lounès Chikhi, Josué Corujo Rodr\'iguez, Simona Grusea, Max Halford, Marine Ha-Shan, Alexane Jouniaux, Inês Lourenço, Beatriz Mourato, Khoa N'Guyen, Cyriel Paris, Didier Pinchon, Willy Rodr\'iguez Valcarce, Jordi Salmona, Patricia Santos, Rémi Tournebize... But it should also be stressed that these results would not have been possible without the direct or indirect collaboration of a multitude of researchers that it is impossible to mention here, and of which Camille Noûs is the representative.}


\bibliographystyle{apalike}
\bibliography{references}
\end{document}